\documentclass[preprint,aps, amsfonts, nofootinbib]{revtex4}

\usepackage{epsf}
\usepackage{amscd}
\usepackage{amsmath}
\input xy
\xyoption{all}

\newcommand{\beq}{\begin{equation}}
\newcommand{\eeq}{\end{equation}}
\newcommand{\bea}{\begin{eqnarray}}
\newcommand{\eea}{\end{eqnarray}}

\newcommand{\nn}{\nonumber}
\newcommand{\benn}{\begin{displaymath}}
\newcommand{\eenn}{\end{displaymath}}
\newcommand{\ket}[1]{| #1 \rangle}                     %  | >
\newcommand{\bra}[1]{\langle #1 \, |}                  %  < |
\newcommand{\Lag}{{\mathcal L}}

\newcommand{\qv}{{\mathbf{q}}}
\newcommand{\pv}{{\mathbf{p}}}
\newcommand{\Pv}{{\mathbf{P}}}
\newcommand{\Dv}{{\mathbf{D}}}

\def\slashchar#1{\ensuremath{                               %
   \setbox0=\hbox{${}#1{}$}       % set a box for #1 
   \dimen0=\wd0                                 % and get its size
   \setbox1=\hbox{/} \dimen1=\wd1               % get size of /
   \ifdim\dimen0>\dimen1                        % #1 is bigger
      \rlap{\hbox to \dimen0{\hfil/\hfil}}      % so center / in box
      {}#1{}                                    % and print #1
   \else                                        % / is bigger
      \rlap{\hbox to \dimen1{\hfil${}#1{}$\hfil}}   % so center #1
      /                                         % and print /
   \fi}}                                        %

\begin{document}

\title{Twisted valence quarks and hadron interactions on the lattice} 
\author{Paulo
  F.~Bedaque\footnote{{\tt pfbedaque@lbl.gov}}} \affiliation{Lawrence-Berkeley
  Laboratory, Berkeley, CA 94720, USA} 
\author{Jiunn-Wei Chen\footnote{{\tt jwc@phys.ntu.edu.tw}}} \affiliation{Department of Physics and National Center for Theoretical Sciences at Taipei, National Taiwan University, Taipei, Taiwan 10617}
\preprint{LBNL-56738}
\begin{abstract}
 We consider QCD with valence and sea quarks obeying different boundary conditions. We point out that the energy of low lying  two hadron states do not depend on the boundary condition of the sea quarks (up to exponentially small corrections). Thus, the advantages in using twisted boundary conditions on the lattice QCD extraction of nucleon-nucleon  phase shifts can be gained without the need of new gauge configurations, even in fully unquenched calculations.
\end{abstract}
\maketitle
Particle scattering is defined at infinite volumes.
Hadron-hadron interactions in lattice QCD can be studied, paradoxically,  only at finite volumes. This is because {\it euclidean} correlators with two hadrons are dominated, at large euclidean times, by the state formed by the two hadrons {\it at rest}.  The exponential decay in imaginary time is set by the sum of the two hadron masses and no scattering information can be deduced from the correlator. This observation was formalized in \cite{maiani_testa}. At finite volumes though, the hadrons are forced to interact and the  energy levels depend on the their interaction. Thus, measuring the energy levels, one can deduce information about their interaction \cite{hamber_et_al, luscher_1, luscher_2}. An elegant formula relating the energy levels to the phase shifts was given in \cite{luscher_1} and we will call this approach the ``Luscher method".

One inconvenience with the Luscher method is that the phase shifts are obtained only at the discrete values of the eigenstates energies. For small boxes these values are very separated from each other. One can, of course, change the volume of the box and learn about the phase shifts at other energy values but this is, frequently, prohibitively expensive.
In applying this method to the two-nucleon system a related problem arises. The unnaturally large value of their s-wave scattering length, that is, the strength of their interaction, shifts the energy levels far away from their non-interacting values. For box sizes smaller than about $8$ fm the energy levels are not in the region described by effective range theory and appear at {\it negative} values of the energy which, from the point of view of scattering, are unphysical \cite{beane_et_al}. The large lattice sizes needed to extract effective range parameters values make the prospect of studying nuclear interaction through lattice QCD even more distant. 

In \cite{bedaque_ab} it was suggested that the use of twisted boundary conditions for the quarks or, which is the same, simulations done under the presence of a constant background magnetic {\it potential} coupled to baryon number, shifts the energy levels in a calculable manner (see also \cite{tantalo}). This allows for  an extra handle on the values at which the phase shifts can be determined. This method, however,  requires the generation of new gauge configurations at each different value of the background field used. The purpose of this report is to point out that, under some circumstances, the same effect can be obtained by coupling only the {\it valence} quarks to the background field, which obviates the need of new gauge configurations generated with a background field. We also discuss how and when this method can be used in meson-baryon and meson-meson systems. As this paper was being finished, a similar  point was made independently by Sachrajda and Villadoro \cite{sachrajda}, in the context of pion physics.

%%%%%%%%%%%%%%%%%%%%%%%%%%%%%%%%%%%%%%%%%%%%%
Let us consider QCD with the sea quarks satisfy standard periodic boundary conditions while the valence quarks have ``twisted" conditions:

\bea
q_v(x,y,L) &=& e^{i \theta^a \tau^a}q_v(x,y,0),\nn\\
q_s(x,y,L) &=& q_s(x,y,0),
\eea where $q_v$ are the valence quarks, $q_s$ the sea quarks and $\tau^a = (1,\vec{\tau})$ acts on isospin space. 
This is the theory of partially twisted QCD (ptQCD) studied independently in \cite{sachrajda}. A field theoretical description of ptQCD can be given using the trick used in partially quenched QCD \cite{morel}. 
We write the partition functions as 

\beq\label{Zbc}
Z = \int_{q_v(L) = e^{i \bf\theta . \bf\tau}q_v(0)}  Dq_v D\bar q_v Dq_s D\bar q_s D\tilde q D\tilde\bar q e^{-\int d^4x \left[ \frac{1}{4}F^2+\bar q_v(\slashchar{D}+m)q_v + \bar{\tilde q} (\slashchar{D}+m)\tilde q + \bar q_s (\slashchar{D}+m)q_s\right]},
\eeq where $\tilde q$ are ``ghost" (bosonic) quarks satisfying the same boundary conditions as the valence quarks.  Integration over the valence and ghost quark fields gives two determinants that cancel each other and only sea quarks appear on internal quark loops. The ghosts violate the spin-statistics theorem and  ptQCD is not a unitary theory. The ``wrong" statistics for the ghosts will generate  a Hilbert space with a non-positive metric.  Notice that the isospin (and baryon number) of valence, sea and ghost quarks is separately conserved.

We can trade the twisted boundary conditions by the presence of a constant gauge field potential by performing a field redefinition  similar to  a gauge transformation, except for being discontinuous at $z=L$:

\bea
q_v(x,y,z)&\to& e^{i \frac{z\theta^a}{L} \tau^a} q_v(x,y,z),\nn\\
\tilde q(x,y,z)&\to& e^{i \frac{z\theta^a}{L} \tau^a} \tilde q(x,y,z).\\
\eea The effect of this transformation is to change the lagrangian in Eq.(\ref{Zbc}) to

\beq\label{Lptqcd}
\Lag_q = \bar q_v (\slashchar{D}+i\slashchar{A}+m)q_v + \bar{\tilde q} (\slashchar{D}+i\slashchar{A}+m)\tilde q
 + \bar q_s (\slashchar{D}+m)q_s,
\eeq with $A_\mu=(0, \theta^a \tau^a/L \hat{z})$. Obviously, this lagrangian has now  a $SU_{isospin}(2)\times U_B(1)$ gauge symmetry of the form

\bea\label{gauge}
q_v&\to& U(x) q_v, \qquad U(x)\in SU(2)\times U(1)\nn\\
\tilde q_v&\to& U(x) \tilde q_v\nn\\
q_s&\to&  q_s\nn\\
A_\mu&\to& U(x) A U^\dagger(x) -i U \partial_\mu U^\dagger(x).
\eea 
 The maximum value of the external field $A$ is $A=\pi/L$ which, in all interesting cases, is of the order of $m_\pi$ or less. 
Larger values of $A$ can be eliminated by a (continuous) gauge transformation and has no physical consequence.

The sea sector of the theory is identical to QCD since the valence and ghost determinants cancel and  there is no back-reaction of the dynamics of the valence quarks on dynamics of the sea quarks. 
That means that the condensate $\langle \bar q_s q_s\rangle$ is the same as in QCD. 
Due to the symmetries of the theory $\langle \bar q_s q_s\rangle=\langle \bar q_v q_v\rangle$, 
as discussed by \cite{nophi0}.
The whole pattern of symmetry breaking is described by the diagram below:
\bigskip 
 
\xymatrix@C=50pt{
 {SU_L(2N|N) \times  SU_R(2N|N)\times U_{L+R}(1)}
  \ar@{->}[d]_{m\neq 0}
  \ar@{=>}[r]_-{\langle\bar q_v q_v\rangle=\langle\bar q_s q_s\rangle\neq 0}
&{SU_{L+R}(2N|N)\times U_{L+R}(1)} \ar@{-}[2,0]_{A\neq 0} \\
{SU_{L+R}(2N|N)\times U_{L+R}(1) }\ar@{->}[d]_{A\neq 0}&{}\ar@{->}[d]\\
 {SU_{L+R}(N|N)\times SU_{L+R}(N)\times U(1) \times U(1)} 
  \ar@{=>}[r]_-{\langle\bar q_v q_v\rangle=\langle\bar q_s q_s\rangle\neq 0}
&{SU_{L+R}(N|N)\times SU_{L+R}(N)\times U(1) \times U(1)}  }
\bigskip 
 
In the upper left corner we have the full symmetry in the absence of either quark masses or external field. On the lower right corner the remaining symmetry after spontaneous chiral symmetry breaking, a common quark mass for all quarks and the external field are included.
Double arrows indicate spontaneous breaking, single arrows the explicit breaking of symmetries.

This pattern of symmetry breaking implies the existence of Goldstone bosons made up of valence, sea and ghost quarks (in addition to the ones formed only by valence quarks) forming multiplets of the remaining symmetry $SU_{L+R}(N|N)\times SU_{L+R}(N)\times U(1) \times U(1)$. In the $N=2$ case we have not only pions made of $u$ and $d$ quarks but also ghost pions made of $\tilde u$ and $\tilde d$ ghosts, sea pions made of $u_s$ and $d_s$, and mixed combinations including fermionic states like, for instance, $\tilde u \gamma_5 \bar d, \cdots$. A similar phenomena occurs with other hadrons. In hadronic low energy effective theories, hadrons containing sea and ghost quarks occur inside loops and serve the purpose of canceling the contribution of valence quark loops and substituting them by the correct ones with sea quarks instead. The construction described above is very similar to the one used in partially quenched QCD for the meson \cite{bernard_golterman, golterman_leung, sharpe_enhanced, nophi0}, one baryon \cite{chen_savage_partially_n, beane_savage_partially_n} and two-baryon \cite{beane_savage_partially_nn} sectors. An alternative (and equivalent) formalism is given by the replica method \cite{damgaard_replica, damgaard_splittorff}.

%%%%%%%%%%%%%%%%%%% SUBSECTION %%%%%%%%%%%%%%%%%%%%%%%%%%%%

\subsection{Nucleon-nucleon scattering}
Since the number (and isospin) of valence quarks is conserved, all intermediate states in two-nucleon processes contain the two initial baryons.
For the sake of argument we first consider the regime $Q\ll m_\pi$, where $Q$ is the typical momentum of the state considered and $m_\pi$ the pion mass.  The effective theory in this regime  is described by

\bea\label{L_pionless}
\mathcal{L}_{NN} &=& N^\dagger(iD_0 + \frac{\Dv^2}{2M})N 
-{C_{0}}\left( N^{T}P_{i}N\right) ^{\dagger }\left(
N^{T}P_{i}N\right) \ +\ {\frac{1}{8}C_{2}}\left[ (N^{T}P_{i}N)^{\dagger
}\left( N^{T}{\cal O}_{i}N\right) +h.c.\right] +\cdots\nn\\
&+& \tilde N^\dagger(iD_0 + \frac{\Dv^2}{2M})\tilde N 
-C_0(\tilde N^T P_i \tilde N)^\dagger \tilde N^T P_i \tilde N+\cdots\nn\\
&+&\cdots
\eea where the dots represent higher derivative terms, $N$ is the valence nucleon field, $\tilde N$ the field of a nucleon containing one sea quark and
$P_{i}$ is a spin-isospin projector and ${\cal O}_{i}=P_{i}\overrightarrow{{\bf D}}^{2}+\overleftarrow{{\bf D}}%
^{2}P_{i}-2\overleftarrow{{\bf D}}P_{i}\overrightarrow{{\bf D}}$.
Also not shown explicitly are the terms containing the nucleons with other combinations of valence, sea and ghost quarks. 
The background field can only appear within covariant derivatives $\Dv N=D N + i A N$  since the underlying theory has a gauge symmetry (Eq.~\ref{gauge}). 
There are two important points to notice. The first is that the conservation of valence isospin and quark number forbids terms coupling valence and non-valence quarks so the valence nucleons effectively decouple from its analogues containing ghost and/or sea quarks. The second one is that the coefficients in Eq.~\ref{L_pionless} encapsulate physics in the scale $\sim1/m_\pi$ and have only an exponentially small dependence on the volume and boundary conditions of order $e^{-m_\pi L}$.

%%%%%%%%%%%%%%%%%%%%%%%%%%%%%%%%%%
\bigskip
\begin{figure}[!htbp]
  \centerline{{\epsfxsize=3.0in \epsfbox{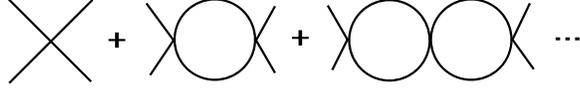}}}
%\vskip 0.15in
\noindent
\caption{Graphs contributing to nucleon-nucleon scattering in the $Q\ll m_\pi$ regime.}
\label{fig:bubbles}
\end{figure}  
%%%%%%%%%%%%%%%%%%%%%%%%%%%%%%%%%%%%%%%%%%%%%
We can obtain the Luscher formula by comparing two calculations in this effective theory, the scattering amplitude in infinite volume and a finite volume two-nucleon correlator. They are both given by the iteration of the same short distance interaction, as shown on Fig. \ref{fig:bubbles}.  The correlator can be computed {\it exactly} (using dimensional regularization), to all orders in the low-energy expansion, as shown below:

\begin{eqnarray}\label{corr_NN}
C(E,\pv) &=&\int dt e^{iEt}\bra{0}T N^T(\pv+{\bf p_0},t) \Pv N(-\pv+{\bf p_0},t) \ .\  N^\dagger(-\pv+{\bf p_0},0) \Pv^{\dagger} N^*(\pv+{\bf p_0},0) \ket{0}\nn\\
&\sim&\mathop{\sum}_{\vec{q}} \frac{i}{E-E_n+i0} |\bra{E_n}N^\dagger(-\pv+{\bf p_0},0) \Pv^{\dagger} N^*(\pv+{\bf p_0},0) \ket{0}|^2\nn\\
&\sim& \frac{1}{1-\mathop{\sum}_{n} C_{2n} (ME)^{n}\frac{1}{L^3}\mathop{\sum}_{\qv} \frac{1}{E-\frac{q^2}{M}}} ,
\end{eqnarray} 
 where the sum is over the allowed momenta 
\begin{eqnarray}\label{allowed_q}
q_{x,y} &=& \frac{2\pi}{L} n_{x,y},\nonumber\\
q_z &=& \frac{2\pi}{L} (n_z+\frac{\phi}{2\pi}),
\end{eqnarray} with $n_x, n_y$ and $n_z$ integers, and ${\bf p_0}=\phi/L \hat{z}$  is the minimum momentum allowed in the lattice.

By  the other hand the same combination of low energy constants appearing in Eq.(\ref{corr_NN}) also determines the scattering in the infinite volume limit

\begin{eqnarray}\label{amplitude}
{\mathcal T} &=&\frac{\sum_n C_{2n}(\mu) k^{2n}}{1-I_0\sum_n C_{2n}(\mu) k^{2n} }\nonumber\\
&=&\frac{4\pi}{M}\frac{1}{k\cot \delta -ik} .
\end{eqnarray} Comparing Eqs.~\ref{corr_NN} and \ref{amplitude} we can relate the shift in the pole of the correlator in Eq.(\ref{corr_NN}) to the infinite volume phase shifts. This provides the generalization of the Luscher formula to the twisted boundary condition case, as discussed in \cite{bedaque_ab}.

This shows that in ptQCD, baryon-baryon scattering at very low energies is the same as in QCD with twisted (valence {\it and} sea) quarks.
The numerical estimates of the influence of the boundary conditions on the two-nucleon states is then unchanged from \cite{bedaque_ab}.

\begin{figure}[!htbp]
  \centerline{{\epsfxsize=2.0in \epsfbox{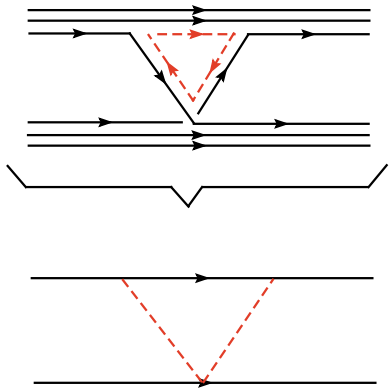}}}
%\vskip 0.15in
\noindent
\caption{Contribution to nucleon-nucleon scattering involving sea quarks.}
\label{fig:2pionX}
\end{figure}

This result  is valid up to a higher energy regime. In fact, it is correct up to the threshold for pion production.
To show this we need to modify the derivation given above.
At higher momenta of order $Q\sim m_\pi$ the appropriate effective theory contains mesons explicitly. Valence baryons can then couple to non-valence baryons through the exchange of non-valence mesons. We should distinguish two kinds of scattering  graphs: irreducible ones (which do not come apart by cutting two valence nucleon lines) and reducible ones. 
The non-valence nucleons and mesons appear inside  the irreducible parts of diagrams only (see e.g. Fig.~\ref{fig:2pionX}),  the finite volume effects of which are of order $e^{-m_\pi L}$. It remains true that intermediate states, sensitive to finite volume effects,  are always composed of two valence nucleons. The full amplitude results from iterating the irreducible graphs or, what is the same, solving the Lippman-Schwinger equation:

\beq\label{ls}
\mathcal{T}(k,p) = V(k,p)-\int \frac{d^3q}{(2\pi)^3}  \mathcal{T}(k,q)\frac{1}{E-\frac{q^2}{M}+i0} V(q,p),
\eeq where $V(k,p)$ is the sum of the irreducible graphs (potential). The scattering amplitude is related to the $\mathcal{K}$-matrix satisfying
\beq\label{kequation}
\mathcal{K}(k,p) = V(k,p)-\int \frac{d^3q}{(2\pi)^3}  \mathcal{K}(k,q)\mathcal{P}\left(\frac{1}{E-\frac{q^2}{M}}\right) V(q,p)
\eeq through

\beq\label{TandK}
\mathcal{T}(k,p) = \mathcal{K}(k,p) \left(  1+\frac{i Mk}{4\pi}\mathcal{T}(k,k)\right).
\eeq The $\mathcal{K}$-matrix is related to the phase shifts by $1/\mathcal{K}(k,k)=Mk\cot\delta/4\pi$, as can be seen from Eqs.~\ref{amplitude} and \ref{TandK}. Eq.~\ref{kequation} is modified at finite volume only by the change of the integral over intermediate states by a sum 

\beq
\int \frac{d^3q}{(2\pi)^3}\cdots \rightarrow \frac{1}{L^3}\sum_{\vec{q}}\cdots, 
{\rm with}\quad
q_{x,y} = \frac{2\pi}{L} n_{x,y}, 
q_z = \frac{2\pi}{L} (n_z+\frac{\phi}{2\pi}).
\eeq The typical momentum of this sum is determined by the scale of $V(k,p)$, in our case, $q\sim m_\pi$. As long as the box satisfies $m_\pi L\gg 1$ the effect of the discretization of the intermediate 
momenta is small and the $\mathcal{K}$-matrix at finite volume differs from its infinite volume counterpart by terms proportional to $e^{-m_\pi L}$. For instance, the leading source of boundary condition dependence in nucleon-nucleon scattering is the one-pion exchange part of the nuclear potential, as this is the longest range contribution of the potential. The equation determining the $\mathcal{K}$-matrix at finite volume is (after projecting in the spin singlet)

\begin{eqnarray}
\mathcal{K}(k,p) &=& V(k,p) - \frac{1}{L^3}\sum_{\vec{q}} \mathcal{K}(k,y)\frac{1}{E-\frac{q^2}{M}} \frac{g_{\pi N}^2}{(q-p)^2+m^2}\\
&=& V(k,p) - \frac{1}{L^3}\sum_{\vec{q}}\int d^3y \delta(\vec{q}-\vec{y})\mathcal{K}(k,y) \frac{1}{E-\frac{y^2}{M}} \frac{g_{\pi N}^2}{(y-p)^2+m^2}\\
&=&V(k,p) - \underbrace{\sum_{\vec{l}}\int\frac{d^3y}{(2\pi)^3} e^{iL\vec{l}.\vec{y}-il_z\phi}
\mathcal{K}(k,y) \frac{1}{E-\frac{y^2}{M}} \frac{g_{\pi N}^2}{(y-p)^2+m^2}}
_{
\stackrel{mL\ll 1}{\rightarrow} A e^{-mL}(4+2\cos(\phi))},   \nonumber   \\
\end{eqnarray} where the allowed values of $\vec{q}$ are like in Eq.~(\ref{allowed_q}) and
$\vec{l} =2\pi \vec{n}/L $.
 The factor $4+2\cos(\phi)$ comes from summing $e^{-il_z\phi}$ over the six points in wave vector space with unit length. Notice that this is not true for the $\mathcal{T}$-matrix. In the infinite volume limit the integral in Eq.~\ref{ls} is dominated by $q\sim \sqrt{ME}$ and for these values the integral is not well approximated by the discrete sum. The argument presented above in the $Q\ll m_\pi$ regime can now be applied to the $\sqrt{Mm_\pi}>Q\sim m_\pi$ regime just by changing $\sum_n C_{2n} (ME)^n$ by $1/\mathcal{K}(\sqrt{ME},\sqrt{ME})$.

%%%%%%%%%%%%%%%%%%%%%%%%%%%%%%%%%%
\bigskip
%%%%%%%%%%%%%%%%%%%%%%%%%%%%%%%%%%%%%%%%%%%%%

\subsubsection{Meson-meson and meson-nucleon scattering}

\begin{figure}[!htbp]
  \centerline{{\epsfxsize=3.3in \epsfbox{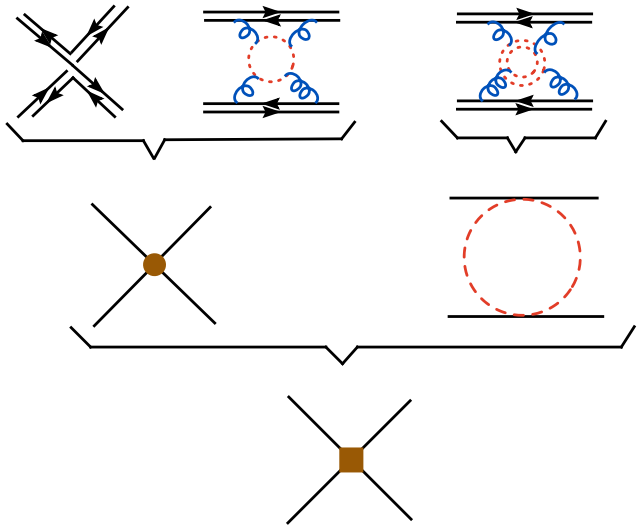}}}
%\vskip 0.15in
\noindent
\caption{Diagrams contributing to $I=2$ $\pi-\pi$ scattering in QCD (top row), in an effective theory ($\chi$PT) valid below the confinement scale (second row) and in a effective theory valid below the pion mass (lower row). Sea quarks and mesons containing sea quarks are denoted by red, dashed lines. }
\label{fig:I2}
\end{figure}

 In systems containing mesons, the presence of valence {\it anti}-quarks may invalidate the point in the previous section. It is important to  distinguish two kinds of  channels. In the first we have those channels where  the initial valence quarks cannot annihilate with each other, for instance,
 the two-pion $I=2$ channel or the $I=3/2$ pion-nucleon channel. In this case, intermediate states have to include the original mesons and sea quarks can appear only on t-channel exchanges (see Fig.(\ref{fig:I2})). 

%%%%%%%%%%%%%%%%%%%%%%%%%%%%%%%%%%
\bigskip
%%%%%%%%%%%%%%%%%%%%%%%%%%%%%%%%%%%%%%%%%%%%%
\noindent
The situation in these channels is identical to the two-nucleon case discussed above. The non-unitarity due to the different boundary conditions for sea and valence quarks is relatively harmless at low energies and manifests itself only
on terms suppressed by $e^{-m_\pi L}$.
 The end result is that the results of ptQCD are the same as the ones in fully twisted QCD, and one can relate their energy levels to QCD phase shifts through the generalization of the Luscher formula discussed in \cite{bedaque_ab}. In the two-pion, $I=2$ channel, the external field needs to couple to the third isospin component (or alternatively, we can use opposite boundary conditions for up and down quarks), if it is to have any effect on the energy levels. This coupling breaks isospin but conserves $I_3$, and preventing  mixing between the $I=2, I_3=2$ state from the $I=0$ state.

%%%%%%%%%%%%%%%%%%%%%%%%%%%%%%%%%%
\bigskip
\begin{figure}[!htbp]
  \centerline{{\epsfxsize=3.3in \epsfbox{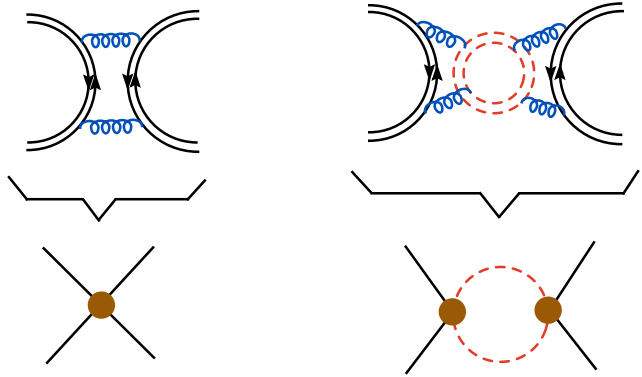}}}
%\vskip 0.15in
\noindent
\caption{Diagrams contributing to $I=0$ $\pi-\pi$ scattering in QCD (top row), in an effective theory ($\chi$PT) valid below the confinement scale (second row) and in a effective theory valid below the pion mass (lower row). Sea quarks and mesons containing sea quarks are denoted by red, dashed lines. }
\label{fig:I0}
\end{figure}  
%%%%%%%%%%%%%%%%%%%%%%%%%%%%%%%%%%%%%%%%%%%%%
 
In systems where the incoming valence quarks can annihilate the situation is more complicated. Intermediate states with only hadrons containing at least one sea/ghost quarks can occur (see Fig.(\ref{fig:I0})). In these diagrams the rest mass of the incoming hadrons is released and  the intermediate states can go on-shell. They generate cuts in the amplitude all the way down to zero momentum and cannot be integrated out. 
Even at low energies the lack of unitarity is evident and there is no way to relate energy levels to phase shifts in an exact way. One can, of course, develop a chiral perturbation theory adequate to partially twisted QCD in the molds of partially quenched chiral perturbation theory, and then relate lattice observables to low energy constants \footnote{These coefficients are the same in ptQCD as in QCD, for the same reasons the low energy constants in partially quenched QCD are the same ones as those of QCD itself.}. This method, however, can only be accurate to a certain order in the chiral expansion and it is unclear if there is any advantage in using partially twisted QCD in these cases.

% As this paper was being finished, reference \cite{sachrajda} appeared. In there, some of the conclusions we arrived here are discussed in the context of pion physics. 
% In particular, it is pointed out the equality between full  and partial twisting in the pion-pion $I=2$ channel as far as low energy scattering is concerned.

 The authors would like to thank conversations with M. Savage and D. Lin. This work was supported in part by the Director, Office of Energy Research,
  Office of High Energy and Nuclear Physics, by the Office of Basic Energy
  Sciences, Division of Nuclear Sciences, of the U.S. Department of Energy
  under Contract No.~DE-AC03-76SF00098  and the National Science Council of ROC.
%%%%%%%%%%%%%%%%%%%%%%%%%%%%%%%%%%
  
\end{document}